\documentclass[twocolumn,aps,showpacs]{revtex4}
\usepackage{amssymb}
\usepackage{amsmath,bm}
\usepackage{graphicx}
\usepackage[normalem]{ulem}
\usepackage{multirow}

\usepackage[usenames, dvipsnames]{xcolor}
\usepackage{tensor}

\renewcommand{\sout}{\bgroup \color{red} \ULdepth=-.5ex \ULset}

\begin{document}

\title{ The QCD critical point from the Nambu-Jona-Lasino model with a scalar-vector  interaction}
\author{Kai-Jia Sun\footnote{%
kjsun@tamu.edu}}
\affiliation{Cyclotron Institute and Department of Physics and Astronomy, Texas A\&M University, College Station, Texas 77843, USA}
\author{Che-Ming Ko\footnote{%
ko@comp.tamu.edu}}
\affiliation{Cyclotron Institute and Department of Physics and Astronomy, Texas A\&M University, College Station, Texas 77843, USA}
\author{Shanshan Cao\footnote{%
shanshan.cao@wayne.edu}}
\affiliation{Cyclotron Institute, Texas A\&M University, College Station, TX, 77843, USA}
\affiliation{Department of Physics and Astronomy, Wayne State University, Detroit, MI, 48201, USA}
\author{Feng Li\footnote{%
fengli@lzu.edu.cn}}
\affiliation{School of Physical Science and Technology, Lanzhou University, Lanzhou, Gansu, 073000, China
}

\date{\today}

\begin{abstract}
We study the critical point in the QCD phase diagram  in the Nambu-Jona-Lasino (NJL) model by including a scalar-vector coupled interaction.  We find that varying the strength of this interaction, which has no effect on the vacuum properties of QCD,  can significantly affect the location of the critical point in the QCD phase diagram, particularly the value of the cirtical temperature.  This provides a convenient way to use the NJL-based transport or hydrodynamic model to extract  information about the QCD phase diagram from relativistic heavy-ion collisions.
\end{abstract}

\maketitle

\section{Introduction}\label{introduction}

Studying the QCD phase structure is among the most important goals of ongoing experiments on heavy-ion collisions~\cite{Braun-Munzinger:2015hba,Luo:2017faz,Bzdak:2019pkr}. By changing the beam energy and selecting different system sizes and the rapidities of measured particles, it is possible to probe different regions of the QCD phase diagram, particularly  the critical  endpoint (CEP)~\cite{Stephanov:2007fk} on the first-order phase transition line. To make this possible  requires, however, versatile dynamic models  to describe the expansion of  created hot dense matter with a flexible equation of state that can have the critical point at varying temperatures ($T$) and baryon chemical potentials ($\mu_B$) in the QCD phase diagram~\cite{Nahrgang:2015tva,Bluhm:2020mpc,Murase:2016rhl,Hirano:2018diu,Nahrgang:2017oqp,Bluhm:2018plm,Singh:2018dpk,Akamatsu:2016llw,An:2019osr,Stephanov:2017ghc,Rajagopal:2019xwg}.
  
At zero and small baryon chemical potentials, the lattice quantum chromodynamics (LQCD)~\cite{Bernard:2004je,Aok06,Baz12} has shown that the  quark-gluon plasma (QGP) to hadronic matter  phase transition is a smooth crossover.  Although it is not yet possible  for the lattice QCD to study the quark-hadron phase transition at high baryon chemical potential  due to the fermion sign problem. Studies based on effective theories  have suggested that the phase transition is a first-order one at large baryon chemical potential~\cite{Asa89,Fuk08,Car10,Bra12,Stephanov:2004wx,Stephanov:2007fk,Fukushima:2013rx,Baym:2017whm}, indicating the existence of a  CEP  on the first-order phase transition line in the $\mu_B-T$ plane of the QCD phase diagram. However, its predicted location  has large uncertainties and even its existence remains unclear~\cite{Stephanov:2004wx}.

Among the effective models for studying the QCD phase diagram at finite baryon chemical potential, a frequently used one is the Nambu-Jona-Lasinio model~\cite{Nam611,Nam612}.  Formulated in terms of quark degrees of freedom~\cite{Eguchi:1976iz,Kikkawa:1976fe}, this model allows the description of chiral phase transition at both finite temperature and chemical potential~\cite{Buballa:2003qv} besides providing a framework to describe hadronic systems in the vacuum based on dynamical chiral symmetry breaking and its restoration~\cite{Vogl:1991qt,Klevansky:1992qe,Hatsuda:1994pi}. The improved NJL model  with the Polyakov-loop (PNJL)  also makes it possible to describe the confinement-deconfinement phase transition of the quark matter~\cite{Meisinger:1995ih,Meisinger:2001cq,Meisinger:2003id,Fukushima:2003fw,Mocsy:2003qw,Ratti:2005jh,Roessner:2006xn,Fuk08}. The parameters in the NJL model and the PNJL model are largely constrained by the vacuum properties of QCD and the known chiral  dynamics in  hadron  systems at zero temperature. The predicted temperature of the critical point in the NJL model varies from 40 to 80 MeV~\cite{Stephanov:2004wx, Buballa:2003qv}, while its value in the PNJL model can be larger than 100 MeV~\cite{Roessner:2006xn,Costa:2010zw}. For the purpose of locating the critical point via comparing model calculations  with the experimental data  from heavy-ion collisions, it will be useful to extend  the NJL-type models to further expand the region  in the $\mu_B-T$ plane where possible locations of the critical point can be accommodated.

Although a repulsive vector interaction can be included in the NJL or PNJL model to change the critical temperature  of the chiral and/or deconfinement phase transition~\cite{Fuk08},  it, however, leads to a decrease of the critical temperature, making the deviation from the LQCD results even larger~\cite{Steinheimer:2014kka}. Another way  to extend the NJL model is to include  higher-order multi-quark interactions. Besides the  six-quark interaction term from the 't Hooft determinant interaction that breaks the $U_A(1)$ symmetry~\cite{Hof78}, the eight-quark interactions including scalar-scalar, vector-vector, and scalar-vector coupled interaction terms have also been considered~\cite{Kashiwa:2006rc,Osipov:2005tq,Osipov:2006xa}. These higher-order interactions are produced from quantum effects in the high momentum region of the nonperturbative renormalization group calculation~\cite{Aoki:1999dw}. Since the attractive scalar-scalar coupled interaction affects the QCD vacuum properties, its strength is constrained and  can not be arbitrarily changed to modify the location of the critical point.  Although the repulsive vector-vector coupled interaction does not affect the QCD vacuum properties, it always decreases the critical temperature of baryon-rich quark matter, similar to the effect due to the vector interaction.   For the scalar-vector coupled interaction, it is known to be important for reproducing the nuclear saturation properties when using the NJL-type model for nuclear matter~\cite{Koch:1987py}.  As to its application to the quark-hadron phase transition~\cite{Lee:2012fj}, it turns out to be a good candidate because it has no effects on the QCD vacuum properties, and more importantly, its strength can affect the location of the critical point  as to be shown below. By tuning the coupling constant of the scalar-vector coupled  interaction, we can easily change  the location of the critical point in the phase diagram from low  to very high temperatures. These features of the scalar-vector coupled interaction term were not fully appreciated in previous studies~\cite{Kashiwa:2006rc,Osipov:2005tq,Osipov:2006xa}. 

In the present work, we first calculate the phase diagram from  the two-flavor NJL model  by including the scalar-vector coupled interaction among quarks. We then extend the calculations to the three-flavor case  and also to the PNJL model  to study in detail its effect on the location of the critical point in the QCD phase diagram.

\section{The Scalar-Vector coupled interaction in the (P)NJL model}

\subsection{The two-flavor NJL model}

We begin by considering the two-flavor NJL model that is usually
described by the following Lagrangian density~\cite{Buballa:2003qv},  
\begin{eqnarray}
 \mathcal{L}_\text{NJL}^\text{SU(2)}&=&\mathcal{L}_0 + \mathcal{L}_{\text{S}}+ \mathcal{L}_{\text{SV}} , 
 \label{eq:njl2-Lag}
\end{eqnarray}
with
\begin{eqnarray}
&& \mathcal{L}_0 = \bar{\psi}(i\gamma^\mu \partial_\mu - \hat{m})\psi ,  \label{eq:njl2-Lagx} \notag\\
&& \mathcal{L}_{\text{S}} = G_{S} [(\bar{\psi}\psi)^2 + (\bar{\psi}i \gamma_5 \vec{\tau} \psi)^2], \notag\\
&&\mathcal{L}_{\text{SV}}= G_{SV}[(\bar{\psi}\psi)^2 + (\bar{\psi}i \gamma_5 \vec{\tau} \psi)^2]  \notag\\
&&\qquad\quad\times[(\bar{\psi}\gamma^\mu\psi)^2 + (\bar{\psi} \gamma_5 \gamma^\mu\vec{\tau} \psi)^2].  
 \label{eq:njl2-sv}
\end{eqnarray}
In the above, $\psi = (u,d)^T$ represents the 2-flavor quark fields, $\hat{m}=\text{diag}(m_u,m_d)$ is the current quark mass matrix, $\gamma^\mu$ and $\gamma_5=i\gamma_0\gamma_1\gamma_2\gamma_3$ are Dirac matrices, and $\vec{\tau}=(\tau_1,\tau_2,\tau_3)$ is the Pauli matrices in the flavor space. The Lagrangian densities $\mathcal{L}_0$, $\mathcal{L}_{\text{S}}$, and $\mathcal{L}_{\text{SV}}$ are, respectively, for the free quarks and their scalar and pseudoscalar interactions with the coupling constant $G_S$  as well as the scalar-vector, scalar-axial vector, pseudoscalar-vector and pseudoscalar-axial vector coupled interactions with the coupling constant  $G_{SV}$. We note that the sign of the $G_{SV}$ term in Eq.~(\ref{eq:njl2-sv}) is  the same as the original one introduced in Ref.~\cite{Koch:1987py}, which is opposite to that used in Refs.~\cite{Kashiwa:2006rc,Lee:2012fj}.

As in most studies using the NJL model, we adopt the mean-field approximation~\cite{Wei91} to linearize the model by introducing the following substitutions,
\begin{eqnarray}
(\bar{\psi}\Gamma_i \psi)^2 &=&  2 \bar{\psi} \Gamma_i \psi \langle\bar{\psi} \Gamma_i \psi\rangle - \langle\bar{\psi} \Gamma_i \psi\rangle^2  \notag \\
(\bar{\psi} \Gamma_i \psi \bar{\psi} \Gamma_j \psi)^2 &=&  \langle\bar{\psi}\Gamma_i \psi\rangle^2 (2\bar{\psi}\Gamma_j \psi \langle\bar{\psi}\Gamma_j\psi\rangle)     \notag \\
&&+\langle\bar{\psi}\Gamma_j \psi\rangle^2 (2\bar{\psi}\Gamma_i \psi \langle\bar{\psi}\Gamma_i\psi\rangle) \notag \\
&&- 3\langle\bar{\psi}\Gamma_i \psi\rangle^2 \langle\bar{\psi}\Gamma_j \psi\rangle^2,
\end{eqnarray}
where $\Gamma = \{1,i \gamma_5\vec{\tau},\gamma_\mu,\gamma_5\gamma_\mu\}$ and the angular bracket  
denotes the expectation value from the quantum-statistical average. Due to the
parity symmetry in a static quark matter, one has $\langle\bar{\psi}\gamma^k 
\psi\rangle =\langle\bar{\psi}\gamma_5  \vec{\tau}\psi\rangle= 
\langle\bar\psi\gamma_5\gamma^\mu\psi\rangle=0$, and the Lagrangian density can then be  rewritten as
\begin{eqnarray}\label{NJL}
&&\mathcal{L}_\text{NJL}^\text{SU(2)}= \bar{u}(\gamma^\mu i\partial_{u\mu} - M_u)u+\bar{d}(\gamma^\mu i\partial_{d\mu} - M_d)d\notag\\
&&\quad+2G_{SV}(\rho_u+\rho_d)(\phi_u+\phi_d)^2(\bar{u}\gamma^0 u+\bar{d}\gamma^0 d)\notag \\
&&\quad-G_S(\phi_u +\phi_d)^2 -3G_{SV}(\phi_u +\phi_d)^2(\rho_u+\rho_d),
\end{eqnarray}
In the above, $M_u$ and $M_d$ are the in-medium effective masses of $u$ and $d$ quarks, respectively, given by
\begin{eqnarray}
M_{u}&=& m_u - 2G_S(\phi_u+\phi_d)\notag\\
&-&2G_{SV}(\rho_u+\rho_d)^2(\phi_u+\phi_d), \notag\\
M_{d}&=& m_d - 2G_S(\phi_u+\phi_d)\notag\\
&-&2G_{SV}(\rho_u+\rho_d)^2(\phi_u+\phi_d), 
\label{eq:njl2-mass}
\end{eqnarray}
with $\phi_u=\langle\bar{u}u\rangle$ and $\phi_d=\langle\bar{d}d\rangle$  being the  $u$ and $d$ quark condensates, respectively, and  $\rho_u$ and $\rho_d $  denoting the net $u$ and $d$ quark number  densities, respectively.

The thermodynamic properties of a two-favour quark  matter are determined by the partition function 
$\mathcal{Z} = \text{Tr} [\exp[-\beta(\hat{H}-\mu \hat{N})]]$, where $\beta=1/T$ and $\hat{H}$ are, respectively, the inverse of the temperature $T$ and the Hamiltonian operator, and  $\mu$ and $\hat{N}$ are, respectively, the chemical potential and  corresponding conserved charge number operator. The thermodynamic grand potential of the system is then  given by
\begin{eqnarray}\label{Grand2}
&&\Omega_\text{NJL}^\text{SU(2)}=-\frac{1}{\beta V} \text{ln} \mathcal{Z}\notag\\
&&\qquad=G_S(\phi_u +\phi_d)^2 +3G_{SV}(\phi_u +\phi_d)^2(\rho_u+\rho_d)\notag\\
&&\qquad- 2N_c\sum_{i=u,d}\int \frac{\text{d}^3p}{(2\pi)^3} E_i \notag \\
&&\qquad-2T\sum_{i=u,d}\int \frac{\text{d}^3p}{(2\pi)^3}(z^+(E_i)+z^-(E_i)),
\end{eqnarray}
where $V$ is volume of the system, $N_c=3$ is the number of colors, $E_i=(m_i^2+{\bf p}^2)^{1/2}$,  and 
\begin{eqnarray}\label{z}
z^\pm(E_i)&=& N_c\text{ln}[1+\text{e}^{-\beta (E_i\mp\mu_i^*)}].
\end{eqnarray}
with the effective chemical potentials,
\begin{eqnarray}\label{chemical}
\mu_u^*&=&\mu_u+2 G_{SV}(\rho_{u}+\rho_{d})(\phi_u+\phi_d)^2, \notag \\
\mu_d^*&=&\mu_d+2 G_{SV}(\rho_{u}+\rho_{d})(\phi_u+\phi_d)^2. \label{eq:njl2-mu}
\end{eqnarray}

\begin{table}[t]
\caption{Parameters in the two-flavor \protect NJL model~\cite{Ratti:2005jh,Roessner:2006xn}.}
\begin{tabular}{ c c c c c }

  \hline
  \hline
  $\Lambda$ [MeV] & $G_S\Lambda^2$ & $m_{u,d}$ [MeV]&$M_{u,d}$ [MeV] & $\langle\bar{u}u\rangle^{1/3}$ [MeV] \\
  \hline
   651& 2.135 &  5.5& 325.1 &-251.3\\
   
  \hline
  \hline
\end{tabular}
\label{tab:njl2}
\end{table}

\begin{figure}[t]
  \centering
  \includegraphics[width=0.45\textwidth]{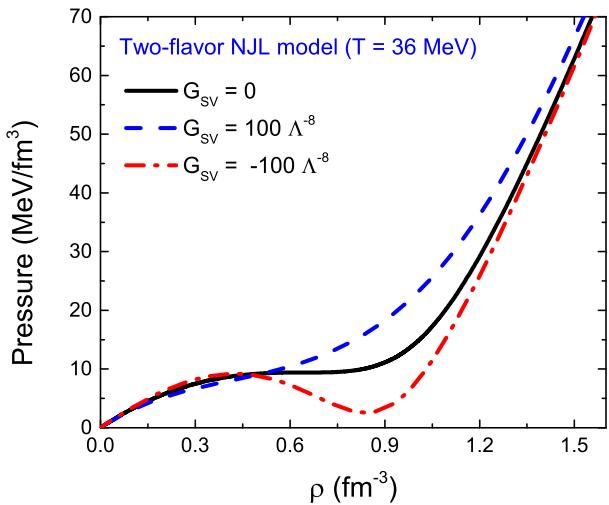}
  \caption{\protect (Color online)  Pressure as a function of  net quark number density at temperature $T=36$~MeV from the two-flavor NJL model  for different values of the scalar-vector coupling constant $G_{SV}$  and with the values of other parameters  given in Table~\ref{tab:njl2}. }
  \label{pic:pressure}
\end{figure}

\begin{figure}[t]
  \centering
  \includegraphics[width=0.45\textwidth]{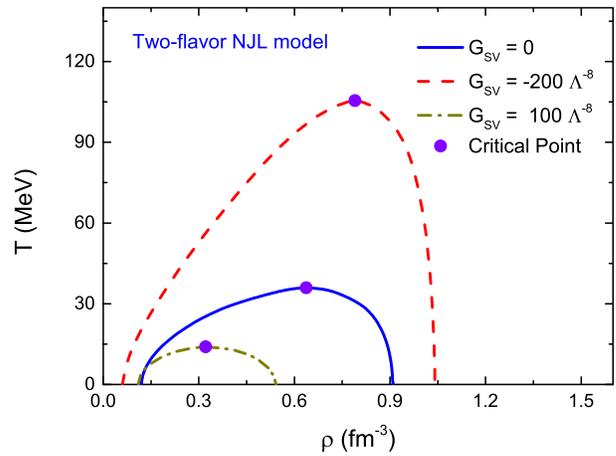}
  \caption{\protect (Color online)  Coexistence lines in the temperature and net quark number density plane from the two-flavor NJL model  for different values of the scalar-vector coupling constant $G_{SV}$  and with the values of other parameters  given in Table~\ref{tab:njl2}. The solid circles denote  corresponding critical points.}
  \label{pic1}
\end{figure}

The quark condensate $\phi_i$ and the net quark number density $\rho_i$ can be determined by minimizing the grand potential, i.e., 
\begin{eqnarray}
\frac{\partial\Omega_\text{NJL}^\text{SU(2)}}{\partial \phi_i}  = \frac{\partial\Omega_\text{NJL}^\text{SU(2)}}{\partial \rho_i}  =0,
\end{eqnarray}
and they are
\begin{eqnarray}
\phi_i &=& 2N_c \int \frac{d^3p}{(2\pi)^3} \frac{M_i}{E_i} (n^{+}_i+ n^{-}_i-1) ,\label{condensate} \\
\rho_i &=& 2N_c \int \frac{d^3p}{(2\pi)^3}(n^{+}_i-n^{-}_i), \label{density}
\end{eqnarray}
with $n^\pm_i=[e^{\beta(E_i\mp\mu_i^*)}+1]^{-1}$.  Because the NJL model is unrenormalizable, a momentum cutoff $\Lambda$ is  needed in evaluating the momentum integral in Eqs.~(\ref{condensate}) and (\ref{density}). In the present study, we employ the parameters $m_u=m_d=5.5$~MeV, $G_S\Lambda^2=2.135$, and a cut-off $\Lambda=651$~MeV~\cite{Ratti:2005jh,Roessner:2006xn}, which are summarized in Table~\ref{tab:njl2} together with the quark in-medium mass and condensate, to study the QCD phase diagram with various values for $G_{SV}$. 

With the quark condensates and net quark density given in the above, one can see from Eq.~(\ref{eq:njl2-mass}) and Eq.~(\ref{eq:njl2-mu}) that the $G_{SV}$ term  affects the effective masses of quarks  and their effective chemical potentials in a quark matter.  Although its effects depend on the quark condensates, which have negative values and increase with decreasing quark density, they also depend  on the quark density.  As a result, including the $G_{SV}$ term  in the NJL model does not affect its description of QCD vacuum properties at zero baryon density, and treating the value of $G_{SV}$ as a free parameter  allows one to obtain different scenarios for the properties of quark matter. 

The effects of the $G_{SV}$ term can be qualitatively understood for quark matter at low density. According to Eq.(\ref{chemical}), a negative $G_{SV}$ resembles a vector interaction in the NJL model~\cite{Buballa:2003qv}, which induces a repulsive interaction among quarks or anti-quarks and an attractive interaction between quark and anti-quark.  Compared to the scalar coupled term $G_S$ in the NJL model, which reduces the quark masses in a medium because of the reduction of quark condensates, a negative $G_{SV}$ counteracts this effect as can be seen from Eq.(\ref{eq:njl2-mass}). With its quadratic dependence on the quark density, the effect of the $G_{SV}$ term on the quark in-medium masses at low quark densities is, however, significantly reduced with increasing quark density, thus resulting in an effectively attractive interaction  among quarks.  Since the repulsive quark interaction due to a negative $G_{SV}$ in the vector channel turns out to be stronger than the attractive quark interaction in the scalar channel for quark matter at low densities,  the net effect of a negative $G_{SV}$ is repulsive.  In quark matter at  very high densities, where  the chiral symmetry is largely restored and the quark condensates are thus close to zero,   the effects of the $G_{SV}$ term become less important, which  is   different from the usual vector interaction in the NJL model~\cite{Buballa:2003qv} that gets stronger at high densities. For quark matter at    intermediate  densities, the effects of the $G_{SV}$ term are, however, more complex, and whether this   leads to a repulsive or an attractive quark interaction depends on the value of the quark density. For a positive $G_{SV}$, its effects on the properties of quark matter are opposite to those of a negative $G_{SV}$. 

The effects of the $G_{SV}$ term can be quantitatively understood from the pressure of a quark matter, 
which is given by $p=-\Omega_\text{NJL}^\text{SU(2)}$, as a function of the net quark number 
density.  In Fig.~\ref{pic:pressure}, we show the results for quark matter at temperature $T=36$~MeV 
 for different values of the scalar-vector coupling constant $G_{SV}$. 
The temperature $T = 36$~MeV is the critical temperature in the two-flavor NJL 
model for $G_{SV}=0$. It is seen that a positive $G_{SV}=$~100$\Lambda^{-8}$ hardens the equation of state at net quark number density of around 0.9 fm$^{-3}$, while a negative $G_{SV}=$~-100$\Lambda^{-8}$ has the opposite effects. As a result, a negative $G_{SV}$ leads to a higher critical temperature while a positive one leads to a lower critical temperature.

\begin{figure}[b]
  \centering
  \includegraphics[width=0.45\textwidth]{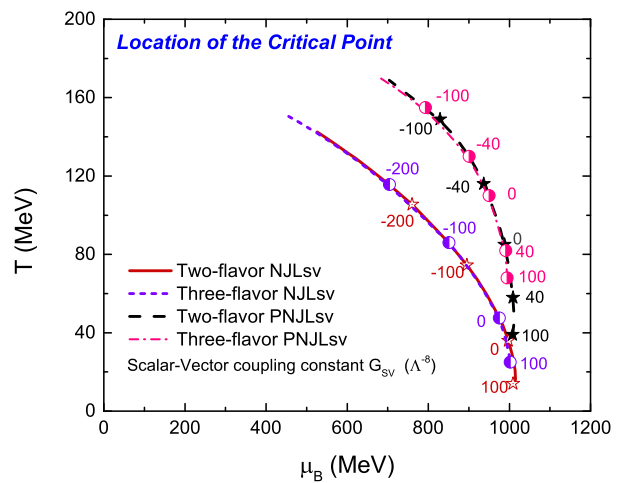}
  \caption{\protect (Color online)  Location of the critical point in two-flavor and three-flavor NJL models and PNJL models with the scalar-vector coupled interaction in the plane of temperature $T$ versus baryon chemical potential $\mu_B$. The lines are obtained by changing the value of the coupling constant $G_{SV}$ with other parameters  in the two-flavor and three-flavor (P)NJL models given in Table~\ref{tab:njl2} and Table~\ref{tab:njl3}, respectively. }
  \label{pic:njlsvcp}
\end{figure}

Figure~\ref{pic1} shows the coexistence line in the temperature and net quark number density plane  for different values of $G_{SV}$. For points on the coexistence line that have same temperature, they  are obtained by requiring equal pressure and chemical potential for the two phases with different quark number densities. The region below the coexistence line is unstable with regard to the phase separation. The blue solid line is the result calculated with $G_{SV}$ = 0, i.e., the default NJL model, and the corresponding critical point is located at temperature $T\approx$ 36 MeV and net quark number density $\rho\approx$  0.64 fm$^{-3}$.  Results obtained with a  scalar-vector coupled interaction of $G_{SV}=$  -200 $\Lambda^{-8}$ are shown by the dashed line, and the critical point in this case shifts to the temperature $T\approx$  105.5 MeV and net quark density $\rho\approx$ 0.79 fm $^{-3}$.  Changing to a scalar-vector coupled interaction of $G_{SV}=$ 100 $\Lambda^{-8}$, reduces the temperature and net quark number density of the critical point  to  $T\approx$ 14 MeV and  $\rho\approx$ 0.32 fm $^{-3}$, respectively. Hence, the critical temperature can be easily increased or decreased by decreasing or increasing the value of $G_{SV}$. For later comparisons of results from the three-flavor NJL model and the NJL model with the Polyakov loop, we also show in Fig.~\ref{pic:njlsvcp} by the red line the locations of the critical point in the temperature and baryon chemical potential plane from the two-flavor NJL model with various values of $G_{SV}$. Although it is  not possible to  make the critical point  approach the $\mu_B=0$ axis by further reducing the value of $G_{SV}$ because its effects on the effective mass and chemical potential vanish on this axis,  the range of values for the critical temperature shown in Fig.~\ref{pic1} by varying $G_{SV}$ is already large enough to cover the region that can be  probed in  realistic heavy-ion collisions.

\subsection{\protect The three-flavor NJL model}

\begin{table}[t]
\caption{Parameters in the three-flavor NJL model~\cite{Buballa:2003qv,Ratti:2005jh,Roessner:2006xn}.}
\begin{tabular}{ c c c c }

  \hline
  \hline
  $\Lambda$ [MeV]~~ $G_S\Lambda^2$ & $K\Lambda^5$ & $m_{u,d}$ [MeV]&$m_{s}$ [MeV]\\

  \hline
   631.4~~ 1.835 & 9.29& 5.5& 135.7\\
  \hline\hline
 $M_{u,d}$ [MeV] & $M_s$ [MeV]& $\langle\bar{u}u\rangle^{1/3}$ [MeV]& $\langle\bar{s}s\rangle^{1/3}$ [MeV] \\ 
  \hline
   335& 527 &-246.9&-267\\  
  \hline
  \hline
\end{tabular}
  \label{tab:njl3}
\end{table}

The three-flavor NJL model includes also the strange quark, which plays an important role in the partonic dynamic of  heavy-ion collisions at high collision energies,  The  Lagrangian density in this model is given by~\cite{Buballa:2003qv}  
\begin{eqnarray}
 \mathcal{L}_\text{NJL}^\text{SU(3)}&=&\mathcal{L}_0 + \mathcal{L}_{\text{S}}+ \mathcal{L}_{\text{SV}} + \mathcal{L}_{\text{det}}, 
\end{eqnarray}
with
\begin{eqnarray}
 \mathcal{L}_0 &=& \bar{\psi}(i\gamma^\mu \partial_\mu - \hat{m})\psi ,  \notag\\
 \mathcal{L}_{\text{S}} &=& G_{S}\sum_{a=0}^8 [(\bar{\psi}\lambda^a\psi)^2 + (\bar{\psi}i \gamma_5 \lambda^a \psi)^2], \notag\\
 \mathcal{L}_{\text{det}} &=& -K[\text{det}\bar{\psi}(1+\gamma_5)\psi + \text{det}\bar{\psi}(1-\gamma_5)\psi], 
\end{eqnarray}
where  $\psi = (u,d,s)^T$ now represents the 3-flavor quark fields and $\hat{m}=\text{diag}(m_u,m_d,m_s)$ is  the corresponding current quark mass matrix.  In the above, $\lambda ^a $($a$=1,...,8)  with $\lambda^0 $ being the identity matrix multiplied by $\sqrt{2/3}$ are the Gell-Mann matrices. The Lagrangian density $\mathcal{L}_{\text{det}}$ is the Kobayashi-Maskawa-t'Hooft (KMT) interaction~\cite{Hof78} that breaks $U(1)_A$ symmetry with `det' denoting the determinant in 
com{the} flavor space~\cite{Bub05}, i.e., $\text{det} (\bar{\psi} \Gamma \psi) = \sum_{i,j,k} (\bar{u}\Gamma q_i)(\bar{d}\Gamma q_j)(\bar{s}\Gamma q_k)$.
This term gives rise to six-point  interactions in three flavors and is responsible for the flavor mixing effect. We assume in the present study that only $u$ and $d$ quarks can have the scalar-vector coupled interaction, so the term $\mathcal{L}_{\text{SV}}$ has the same form as in Eq.~(\ref{eq:njl2-sv}). 

In the mean-field approximation~\cite{Wei91}, the gap equations in the three-flavor NJL model for the quark in-medium effective masses including that ($M_s$) of the strange quark are given by 
\begin{eqnarray}
M_{u}&=& m_u - 4G_S\phi_u + 2K\phi_d\phi_s \notag\\
&&-2G_{SV}(\rho_u+\rho_d)^2(\phi_u+\phi_d), \notag\\
M_{d}&=& m_d - 4G_S\phi_d + 2K\phi_u\phi_s \notag\\
&&-2G_{SV}(\rho_u+\rho_d)^2(\phi_u+\phi_d), \notag\\
M_{s}&=& m_s - 4G_S\phi_s + 2K\phi_u\phi_d. 
\end{eqnarray}
Besides the light quark condensates $\phi_u$ and $\phi_d$ as in the two-flavor NJL model,  there is also the strange quark condensate given by 
\begin{eqnarray}
\phi_s = 2N_c \int \frac{d^3p}{(2\pi)^3} \frac{M_s}{E_s} (n^{+}_s+ n^{-}_s-1),
\end{eqnarray}
where 
$n^\pm_s=[e^{\beta(E_s\mp\mu_s)}+1]^{-1}$ with $E_s=(M_s^2+{\bf p}^2)^{1/2}$ and $\mu_s$ being the the strange quark chemical potential.  The thermodynamic potential of the system can then be written as 
\begin{eqnarray}\label{Grand3}
\Omega_\text{NJL}^\text{SU(3)} &=& 2G_S(\phi^2_u+\phi^2_d+\phi^2_s)\notag\\
&+&3G_{SV}(\phi_u+\phi_d)^2(\rho_u+\rho_d)\notag \\
&-&4K\phi_u\phi_d\phi_s - 2N_c\sum_{i=u,d,s}\int_0^\Lambda \frac{\text{d}^3p}{(2\pi)^3} E_i \notag \\
&-&2T\sum_{i=u,d,s}\int \frac{\text{d}^3p}{(2\pi)^3}(z^+(E_i)+z^-(E_i)).
\end{eqnarray}

To calculate the thermodynamic quantities of a quark matter in the three-flavor NJL model, we employ the parameters $m_u=m_d=5.5$~MeV, $m_s=135.7$~MeV, $G_S\Lambda^2=1.835$, $K\Lambda^5=9.29$, and a cut-off $\Lambda=631.4$~MeV~\cite{Buballa:2003qv}, which are summarized in Table~\ref{tab:njl3} together with the quark in-medium masses and condensates.  The locations of the critical point in the temperature and baryon chemical obtained from the three-flavor NJL model with the scalar-vector coupled interaction are shown in Fig.~\ref{pic:njlsvcp} by the short dashed line.  This line is almost identical to  the solid line from the two-flavor NJL model except that the critical point in the three-flavor case moves to a higher temperature and smaller baryon chemical potential compared to the two-flavor case when the same $G_{SV}$ is used in the two calculations. The main reason for this similarity is  because the parameters in the two-flavor and  three flavor NJL models (see Tables \ref{tab:njl2} and  \ref{tab:njl3}) give similar properties of the QCD vacuum, e.g., the quark condensates and   effective masses in vacuum.  Results from these two models will not be identical if one uses different  values for the parameters~\cite{Buballa:2003qv}. 

In principle, one can also include  the scalar-vector coupled interactions for strange quarks. In this case, the dependence of the critical temperature on the value of $G_{SV}$ becomes much weaker than the results shown in the above, and this is because the in-medium mass of strange quark is much larger than the light quark masses, which makes it  much harder to reach the chiral limit like the light quarks. Since the purpose of present study is to obtain a flexible critical point in the QCD phase diagram by tuning the strength of the scalar-vector coupled interaction,  we have thus neglected the scalar-vector coupled interaction of the strange quark in the  three-flavor NJL and PNJL models. 

\subsection{The NJL model with Polyakov loop}

\begin{figure*}[t]
  \centering
  \includegraphics[width=0.75\textwidth]{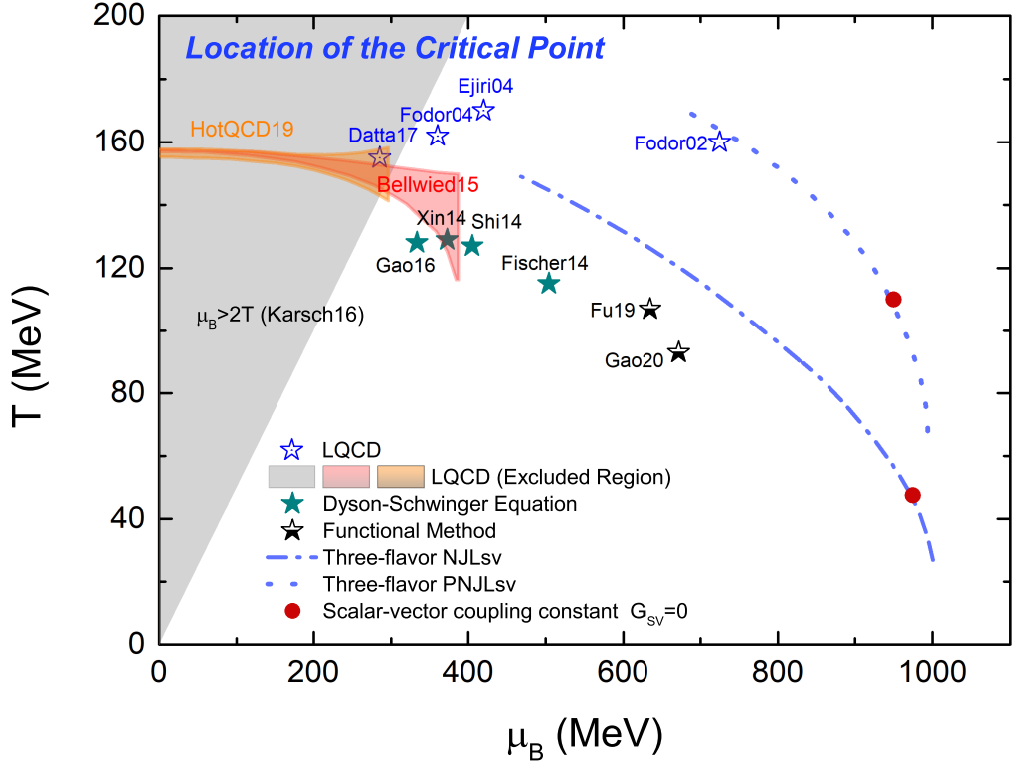}
  \caption{\protect (Color online) \protect Comparison of the location of the critical point in the plane of temperature $T$ versus baryon chemical potential $\mu_B$ from the three-flavor NJL (dash line) and PNJL (dotted line) models by varying the value of the scalar-vector coupling constant $G_{SV}$, with  red solid circles denoting that obtained with $G_{SV}=0$, with  predictions from LQCD~\cite{Ejiri:2003dc,Fodor:2004nz,Datta:2012pj,Karsch:2015nqx,Datta:2016ukp,Bellwied:2015rza,Bazavov:2018mes}, Dyson-Schwinger equation~\cite{Xin:2014ela,Fischer:2014ata,Shi:2014zpa,Gao:2015kea}, and the functional renormalization method~\cite{Fu:2019hdw,Gao:2020qsj}. }
  \label{pic:cp}
\end{figure*}

To  include also the confinement-deconfinement phase transition, a  constant temporal background gauge field representing the Polyakov loops $\Phi$ and $\bar\Phi$ has been added to the NJL model~\cite{Fuk08}. This so-called  PNJL model changes the NJL Lagrangian density to
\begin{eqnarray}
 \mathcal{L}_\text{PNJL}&=&\bar{\psi}(i\gamma^\mu D_\mu - \hat{m})\psi + \mathcal{L}_{\text{S}} +\mathcal{L}_{\text{SV}} \notag \\
  &-& \mathcal{U}(\Phi[A],\bar{\Phi}[A],T), 
 \label{eq:pnjl2-Lag}
\end{eqnarray}
where the covariant derivative is defined as $D^\mu = \partial^\mu-iA^\mu$ with $A^\mu =g\mathcal{A}^\mu_a(x)\lambda_a/2=\delta_0^\mu A_0$ being the $SU(3)$ gluon field in the Polyakov gauge and  $g$ being the QCD strong coupling constant. The effective potential $\mathcal{U}$  for the Polyakov loops is given by 
\begin{eqnarray}
 &&\frac{\mathcal{U}(\Phi,\bar{\Phi},T)}{T^4} = -\frac{1}{2}a(T)\bar{\Phi}\Phi + \notag \\
 && b(T)\text{ln}[1-6\bar{\Phi}\Phi +4(\bar{\Phi}^3+\Phi^3)-3(\bar{\Phi}\Phi)^2], 
 \label{eq:pnjl2-Lagx}
\end{eqnarray}
with 
\begin{eqnarray}
&&a(T) = a_0+a_1\left(\frac{T_0}{T}\right)+a_2\left(\frac{T_0}{T}\right)^2, \notag\\
&&b(T) = b_3\left(\frac{T_0}{T}\right)^3,
\end{eqnarray}
where the parameters $a_0 = 3.51$, $a_1=-2.47$, $a_2=15.2$, and $b_3=-1.75$  are fitted  to the results from the LQCD calculations of the thermodynamic properties of a pure gluon system~\cite{Ratti:2005jh,Roessner:2006xn}. For the temperature parameter $T_0$, its value is  270 MeV,  corresponding to the critical temperature for the deconfinement phase transition of a pure gluon  matter at zero baryon chemical potential~\cite{Fukugita:1989yw}.  The inclusion of quarks leads to a smaller value  of $T_0=210$ MeV. 

The grand potential of a quark matter at finite temperature and quark baryon potential in the PNJL model has a similar expression to Eq.~(\ref{Grand2}) for the two-flavor or Eq.~(\ref{Grand3}) for the three-flavor NJL model except the expression in Eq.~(\ref{z}) is replaced by
\begin{eqnarray}
&&z_\Phi^\pm= \text{ln}[1+3(\bar{\Phi}+\Phi \text{e}^{-\beta (E_i\mp\mu_i^*)})\text{e}^{-\beta (E_i\mp\mu_i^*)}\notag\\
&&\qquad+\text{e}^{-3\beta (E_i\mp\mu_i^*)}].
\end{eqnarray}

As in the NJL model, the quark condensate and quark density are obtained by minimizing the grand potential, i.e., 
$\frac{\partial\Omega_\text{PNJL}^\text{SU(3)}}{\partial \phi_{i}}=\frac{\partial\Omega_\text{PNJL}^\text{SU(3)}}{\partial \rho_{i}}=0$.
Their expressions are similar to those given in Eqs.(\ref{condensate}) and (\ref{density}) except  the color-averaged equilibrium quark  occupation numbers $n_i^\pm$ are replaced by
\begin{eqnarray}
n_\Phi^\pm &=& \frac{\bar{\Phi}\text{e}^{2\beta(E\mp\mu^*)}+2\Phi\text{e}^{\beta(E\mp\mu^*)}+1}{e^{3\beta(E\mp\mu^*)}+3\bar{\Phi}e^{2\beta(E\mp\mu^*)}+3\Phi e^{\beta(E\mp\mu^*)}+1}.\notag\\
\end{eqnarray}
From the above expression, one can see that the quark distribution retains the normal Fermi-Dirac form at high temperature when the Polyakov loops are $\Phi=\bar{\Phi}=1$, while it  becomes the Fermi-Dirac form with a reduced temperature $T/3$ at low temperature when $\Phi=\bar{\Phi}=0$. Hence, the critical temperature in the PNJL model is generally higher than that in the NJL model as the quarks in PNJL model   experience a lower effective temperature.  Note that the PNJL model at zero temperature is identical to the  NJL model. 

Minimizing the grand potential with respect to the Polyakov loops, i.e., $\frac{\partial\Omega_\text{PNJL}^\text{SU(3)}}{\partial \Phi} =\frac{\partial\Omega_\text{PNJL}^\text{SU(3)}}{\partial \bar{\Phi}} = 0$, leads to the following  mean-field equations for $\Phi$ and $\bar{\Phi}$:  
\begin{eqnarray}
&&\partial_\Phi \mathcal{U} \notag\\
&&= 6T\sum_{i=u,d,s}\int \frac{d^3p}{(2\pi)^3} \left[\frac{\text{e}^{-\beta(E_i-\mu_i^*)}}{\exp(z_\Phi^+(E_i))}+\frac{\text{e}^{-2\beta(E_i+\mu_i^*)}}{\exp(z_\Phi^-(E_i))}\right ],\notag \\
&&\partial_{\bar{\Phi}} \mathcal{U} \notag\\
&&= 6T\sum_{i=u,d,s}\int \frac{d^3p}{(2\pi)^3} \left[\frac{\text{e}^{-2\beta(E_i-\mu_i^*)}}{\exp(z_\Phi^+(E_i))}+\frac{\text{e}^{-\beta(E_i+\mu_i^*)}}{\exp(z_\Phi^-(E_i))}\right ]. \notag\\
\end{eqnarray}

In Fig.~\ref{pic:njlsvcp}, we show the locations of the critical point in the plane of temperature and baryon chemical potential obtained from both the two-flavor and the three-flavor PNJL model with the inclusion of the quark scalar-vector coupled interaction.  As shown by the dashed line for the two flavor PNJL model and the dash-dotted line for the three-flavor NJL model, the effects of $G_{SV}$ are similar in these two cases.  We also see that the effect of $G_{SV}$ on the critical chemical potential is  smaller in the PNJL model than in the NJL model for both the two-flavor and the three-flavor case. 

We further compare in Fig.~\ref{pic:cp} the critical point obtained from the three-flavor NJL (dash line) and PNJL (dotted line) models by varying the value of $G_{SV}$, with solid circles denoting those obtained with $G_{SV}=0$, with selected predictions from LQCD~\cite{Ejiri:2003dc,Fodor:2004nz,Datta:2012pj,Karsch:2015nqx,Datta:2016ukp,Bellwied:2015rza,Bazavov:2018mes}, Dyson-Schwinger equation~\cite{Xin:2014ela,Fischer:2014ata,Shi:2014zpa,Gao:2015kea}, and 
the functional renormalization method~\cite{Fu:2019hdw,Gao:2020qsj}.  Predictions from other effective methods can be found in Refs.~\cite{Stephanov:2004wx} and references therein. It is seen that with sufficiently attractive scalar-vector coupled interaction, the locations of the critical point in the NJL and PNJL models can be brought  closer to those predicted  from these first principle approaches.

\section{conclusions}

Based on the NJL model with both two flavors and three flavors as well as with the inclusion of Polyakov loops, we have studied the effect of the eight-fermion scalar-vector coupled interaction, which has no effects on the QCD vacuum properties, on the  critical endpoint of the first-order QCD phase transition line in the QCD phase diagram.   We have found that the location of the critical point in the  temperature and baryon chemical potential plane is extremely sensitive to the strength of this interaction and can be easily shifted by changing its value.  This  flexible dependence of the quark equation of state due to the quark scalar-vector coupled interaction is very useful for locating the phase boundary in QCD phase diagram by  comparing the experimental data  with results from transport model simulations~\cite{Li:2016uvu} or hydrodynamic calculations based on equations of states from such generalized NJL and PNJL models.

\begin{acknowledgments}
One of the authors K. J. Sun would like to thank Dr. Peng-Cheng Chu for helpful discussions. This work was supported in part by the US Department of Energy under Contract No.
DE-SC0015266, the Welch Foundation under Grant No. A-1358,  the U.S. Department of Energy (DOE) under grant number DE-SC0013460, and the National Science Foundation (NSF) under grant number ACI-1550300.
\end{acknowledgments}

\bibliography{myref}
\end{document}